\documentclass[]{aastex7}  

\renewcommand{\baselinestretch}{1.0} 
 
\usepackage{amsmath,amsfonts,amssymb}
\usepackage{graphicx}

\begin{document}

\title{Satellite Streak Brightness Variation with Orbit Height}

\author[0000-0002-2343-0949]{Adam Snyder}
\affiliation{Department of Physics and Astronomy, University of California, Davis, CA 95616}
\email{aksnyder@ucdavis.edu}

\author[0000-0002-9242-8797]{J. Anthony Tyson}
\affiliation{Department of Physics and Astronomy, University of California, Davis, CA 95616}
\email{tyson@physics.ucdavis.edu}

\begin{abstract}
We study the apparent optical brightness of satellite streaks in ground-based observatory cameras as the orbit height is varied. This is done via simulations. Several factors are in play:  the range to the satellite, satellite size and geometry, the diameter of the telescope aperture, and the angular velocity of the satellite image as it streaks across the camera's focal plane.  For a large telescope aperture, a satellite in a lower orbit becomes out-of-focus in a camera focused at infinity.  In the case of Rubin Observatory's LSSTCam, we find that these factors nearly cancel as a large satellite is moved from an orbit at 550 km down to 350 km.
\end{abstract}


\section{Introduction}\label{sec:intro}
The exponential increase in numbers of Low Earth Orbit (LEO) satellites~\citep{falle2023one} presents a challenge to ground-based optical astronomy due to reflected sunlight during much of the night. Historically, attempts to mitigate the optical brightness of these LEO satellites have involved darkening the satellites via various schemes, and exploration of conops changes in collaboration with ground based observatories. Recently, many satellite operators have opted to build larger satellites capable of direct-to-cell communication. At a given reflectivity a larger satellite will scatter more sunlight towards the observatory.   Some observatories image faint galaxies whose surface brightness (flux/pixel) are hundreds of millions times fainter than these LEO satellites. Multiple studies~\citep{walker2020impact,rawls2021satcon2} have thus recommended that satellites be designed such they appear fainter than detectable by the unaided eye in a dark location, 7th V magnitude.  Recent satellites however are being reported as bright as 1st magnitude, a factor of 300 brighter than the recommended brightness. This presents a challenge to worldwide astronomy.

The most impacted observatories are those that have large light collecting apertures, highly sensitive cameras, and wide field of view.   For example, the NSF-DOE Vera C. Rubin observatory will scan the southern sky with 1000 sensitive exposures every night for ten years~\citep{ivezic2019lsst}, the Legacy Survey of Space and Time (LSST). Mitigation of satellite interference with LSST is thus a priority. In this paper we simulate variations in LEO satellite surface brightness observed in the 3200 megapixel camera (LSSTCam) as a function of satellite orbit height.

\section{Satellite Streaks}\label{sec:satellite_streak}

A satellite streak surface brightness profile will depend on the following: bidirectional reflectance distribution function (BRDF), satellite geometric properties, orbital height, and observed angle from zenith. We wish to model the changes to the satellite streak peak brightness due to changes in the orbital height while keeping all other properties the same. We begin with a discussion of the various factors that will affect the measured peak brightness and how these depend on the orbital height.

\subsection{Distance to the Satellite}\label{sec:distance}

\begin{figure}
    \centering
    \includegraphics[width=0.75\linewidth]{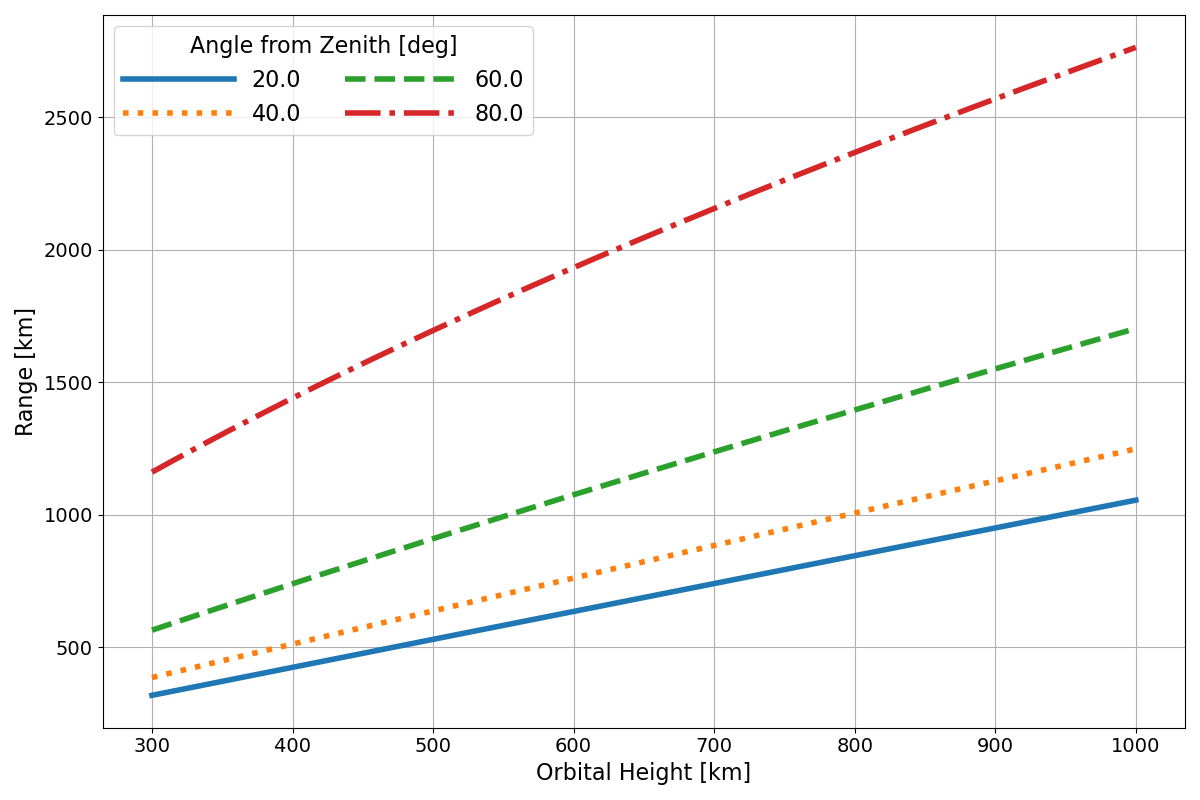}
    \caption{Distance, or range, to satellites observed at specific zenith angles as a function of the orbital height.}
    \label{fig:distance_vs_height}
\end{figure}

The distance from the satellite to the observing telescope, known as the range, is an important quantity because it sets the angular sizes of the satellite as viewed by the telescope and the telescope pupil as viewed from the satellite. For a given zenith angle $\theta_Z$ we define the nadir angle $\theta_N$ as the angle between the nadir of the satellite and a line from the satellite to the telescope, calculated as

\begin{equation}\label{eq:nadir_angle}
    \theta_N = \arcsin\left(\frac{R_\oplus}{R_\oplus + h}\sin\theta_Z\right),
\end{equation}
where $h$ is the orbital height of the satellite. When considering cases of satellites observed at a specific zenith angle, the nadir angle depends only on the orbital height and will simplify a number of calculations. Then the range is

\begin{equation}
    r = \begin{cases} 
            h & \theta_N = 0 \\
            R_\oplus\frac{\sin(\theta_Z - \theta_N)}{\sin\theta_N} & \theta_N \neq 0
        \end{cases}.
\end{equation}
The case of $\theta_N=0$ corresponds to $\theta_Z=0$, i.e. the satellite is observed directly above the telescope. In this case, the range is exactly equal to the orbital height. In general, the range to a satellite observed at a specific zenith angle is more complex and the dependence on orbital height is encapsulated by changes to the nadir angle (see Figure \ref{fig:distance_vs_height}).

\subsection{Brightness}\label{sec:brightness}

The flux reflected from the satellite to the telescope is the fraction of the total incoming flux defined in \cite{fankhauser23} as

\begin{equation}
    \frac{I_\mathrm{out}}{I_\mathrm{in}} = (\hat{w}_i\cdot\hat{n})(\hat{w}_o\cdot\hat{n})f_r(\hat{w}_i,\hat{w_o})\frac{A}{r^2}
\end{equation}

where $\hat{w}_i$ is the vector to the incoming light source, $\hat{w}_o$ is the vector to the telescope, $\hat{n}$ is the normal vector of the satellite surface, $f_r(\hat{w}_i,\hat{w_o})$ is the BRDF, $A$ is the satellite surface area, and $r$ is the range of the satellite. We can then examine each of the terms individually to determine the change in brightness of the satellite. The incoming flux from the Sun $I_\mathrm{in}$ is assumed to be constant since it has a negligible dependence on orbital height. The dot product $\hat{w}_i\cdot\hat{n}$ encapsulates changes in the surface area perpendicular to the source and, for the same reason, is assumed to be unchanged. The dot product $\hat{w}_o\cdot\hat{n} = \cos(\theta_N)$ represents changes in the surface area perpendicular to the observer and depends on the nadir angle defined in Equation \ref{eq:nadir_angle}. Satellites at different orbital heights that are observed at the same zenith angle will have different nadir angles and therefore there will be additional changes to the amount of reflected flux that depends on this projection effect and the specific properties of the satellite BRDF. 

\subsection{Surface Brightness Profiles}\label{sec:sbp}

At a lower orbital height and subsequent shorter range, the angular size of the satellite will increase and it will appear more out-of-focus because the telescope is focused at infinity. The effective FWHM of a satellite streak as a function of range can be approximated as

\begin{equation}
    \theta^2_\mathrm{eff} = \theta^2_\mathrm{atm} + \frac{D^2_\mathrm{satellite} + D^2_\mathrm{mirror}}{r^2},
\end{equation}
where $\theta_\mathrm{atm}$ is the FWHM of a star due to atmospheric turbulence, $D_\mathrm{satellite}$ is the diameter of the satellite (approximated as a circle), and $D_\mathrm{mirror}$ is the diameter of the primary mirror of the telescope \citep{bektesevic18}. From the above approximation we predict that the FWHM of the satellite streak will increase; as a result the surface brightness in electrons per pixel will decrease. 

\subsection{Orbital Velocity}\label{sec:velocity}

Moving objects, such as satellites, are imaged as they traverse the focal plane, whereby individual pixels are exposed to the object for a much shorter time than the full exposure time. We define the effective pixel exposure time $t_\mathrm{eff}$ as the time it takes for the satellite to move the angular distance corresponding to one pixel, which is inversely proportional to the angular velocity at which the satellite appears to move across the focal plane. This angular velocity is related to, but not equal to, the orbital angular velocity; this is because it is measured in the altitude-azimuth coordinate system centered at the telescope rather than in the equatorial coordinate system. 

As in \cite{fankhauser23} this calculation can be simplified by using a satellite-centered frame, although we make some modifications to the axes definitions. As before the z-axis points along the satellite nadir. However for our purposes, the y-axis lies in the plane defined by the center of the Earth, the satellite, and the telescope, rather than the Sun and we apply the restriction that the angle between the y-axis and the vector from the satellite to the telescope be less than 90 deg. Finally, the x-axis is again defined by the right-hand rule.

Important features of this frame-of-reference are that:

\begin{itemize}
    \item Motion of the satellite along the positive y-axis corresponds to an increase in altitude as seen by the telescope.
    \item Motion of the satellite along the positive x-axis corresponds to an increase in azimuth as seen by the telescope.
    \item The vector from the satellite to the telescope is in the yz-frame, at the nadir angle $\theta_N$ that is equivalent to the polar angle of the spherical coordinate representation of this frame-of-reference.
    \item The orbital velocity vector of the satellite is in the xy-frame, at an angle $\phi$ from the x-axis that is equivalent to the azimuthal angle of the spherical coordinate representation of this frame-of-reference.
\end{itemize}

\begin{figure}[htb]
    \centering
    \includegraphics[width=0.75\linewidth]{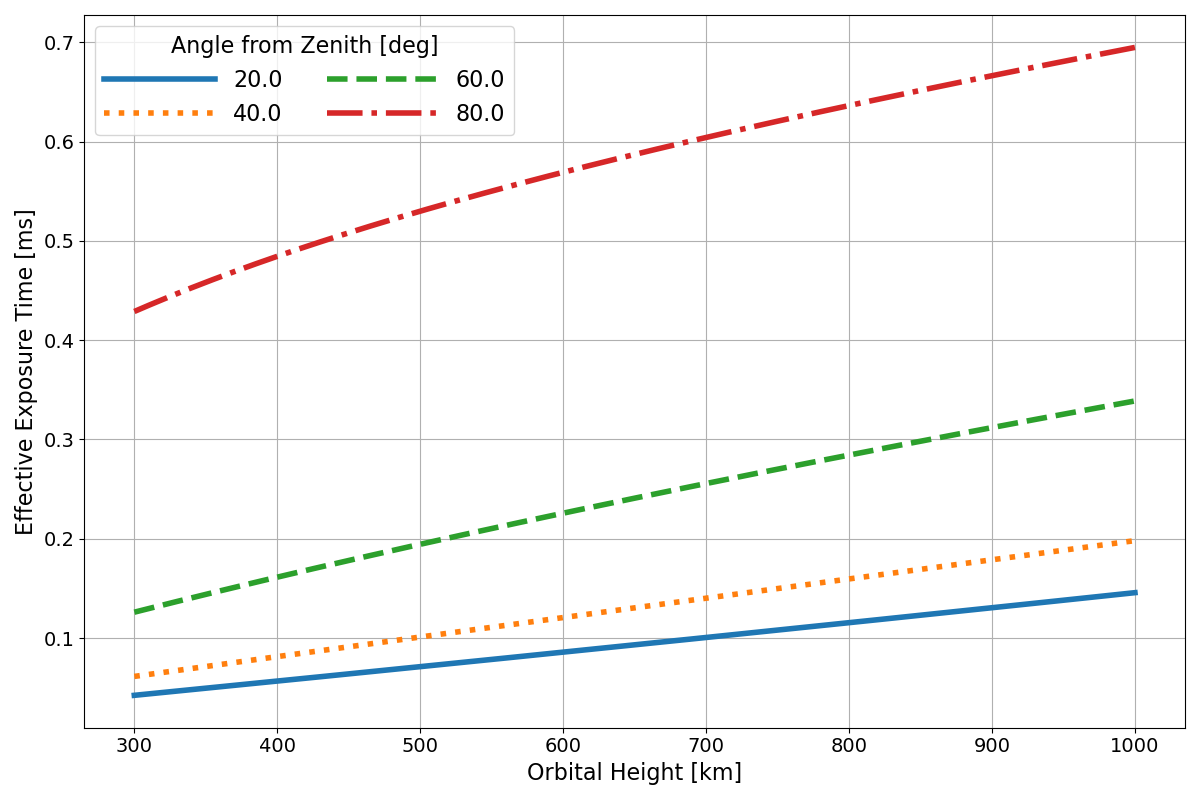}
    \caption{Effective pixel exposure time of satellites observed at specific zenith angles traveling along a constant azimuth as a function of the orbital height.}
    \label{fig:exptime_vs_height}
\end{figure}

The orbital velocity vector $\vec{v}$ has an amplitude of 

\begin{equation}
    v = \sqrt{\frac{GM_\oplus}{R_\oplus + h}}
\end{equation}
that depends on the gravitational constant $G$, the mass of the Earth $M_\oplus$, the radius of the Earth $M_\oplus$, and the orbital height $h$. To calculate the perceived velocity of the satellite in the altitude-azimuth coordinate system we must calculate the component of the orbital velocity vector that is tangential to the vector from the satellite to the telescope as

\begin{equation}
    \vec{v}_T = \vec{v} - (\vec{v}\cdot\hat{r})\hat{r},
\end{equation}
where $\hat{r}$ is the unit vector along the line from the satellite to the telescope. The result can be expressed in terms of the nadir angle $\theta_N$ and the rotation angle $\phi$ as

\begin{equation}
    \vec{v_T} = (v\cos\phi)\hat{x} + (v\sin\phi\cos^2\theta_N)\hat{y} - (v\sin\phi\sin\theta_N\cos\theta_N)\hat{z}.
\end{equation}
To obtain an angular velocity in the altitude-azimuth coordinate system we divide the amplitude of the tangential velocity by the range

\begin{equation}
    \omega_T = \frac{v_T}{r}.
\end{equation}
The result is the tangential angular velocity as measured by the telescope in the altitude-azimuth coordinate system in units of radians per second. It is important to note that this is different than the orbital angular velocity $v/(R_\oplus + h)$, even for cases where $v_T = v$. Finally, to obtain the effective pixel exposure time, we divide the plate scale $p$ of the detector (arcseconds per pixel) by this angular velocity (converted to arcseconds per second),

\begin{equation}
    t_\mathrm{eff} = \frac{p}{\omega_T}.
\end{equation}
For a satellite at lower orbit the orbital velocity term increases and the range decreases, both of which contribute to a decrease in the effective exposure time. Figure \ref{fig:exptime_vs_height} shows the relationship between the effective pixel exposure time and the orbital height of satellites observed at different zenith angles for the case of rotation angle $\phi = \frac{\pi}{2}$. 

An additional contribution to the tangential angular velocity is the effect of the telescope RA/DEC tracking at a rate of 15 deg per hour or $7.27\times10^{-5}$ radians per second. The slowest moving low-earth orbit satellites under consideration are at the highest orbits and are observed at the lowest zenith angles and have tangential angular velocities on the order of $10^{-3}$ radians per second. Therefore this is, even in the most extreme cases, under a 10\% effect. We also note that this effect is independent on the satellite orbital properties and only on the telescope observing scheduler; in this study we are assuming the latter is equal between observations of the satellites at the two simulated orbit heights.

\section{Streak Model Results}\label{sec:results}

In this Section, we simulate a specific case of the difference in the measured peak brightness of streaks produced by satellites at different orbital heights, taking into consideration the various factors detailed in Section \ref{sec:satellite_streak}. The satellite under consideration is modeled as having a 7.25 m by 2.65 m rectangular primary reflecting surface, where the orbital velocity vector is tangential to the axis of the larger dimension. The first step to determining the peak brightness of the satellite streak is to calculate the stationary profile of the satellite, which is the surface brightness profile of the satellite if it were to be tracked by the observing telescope during the observation, following the theory of defocusing described by \cite{bektesevic18}. The stationary profile is the convolution of three components:

\begin{enumerate}
    \item The geometric profile of the satellite determined by the angular size of the satellite as seen from the telescope.
    \item The defocus kernel determined by the angular size of the telescope pupil as seen from the satellite.
    \item The atmospheric PSF represented either as a Kolmogorov or von K\'arm\'an surface brightness profile.
\end{enumerate}

\begin{figure}[htb]
    \centering
    \includegraphics[width=0.75\linewidth]{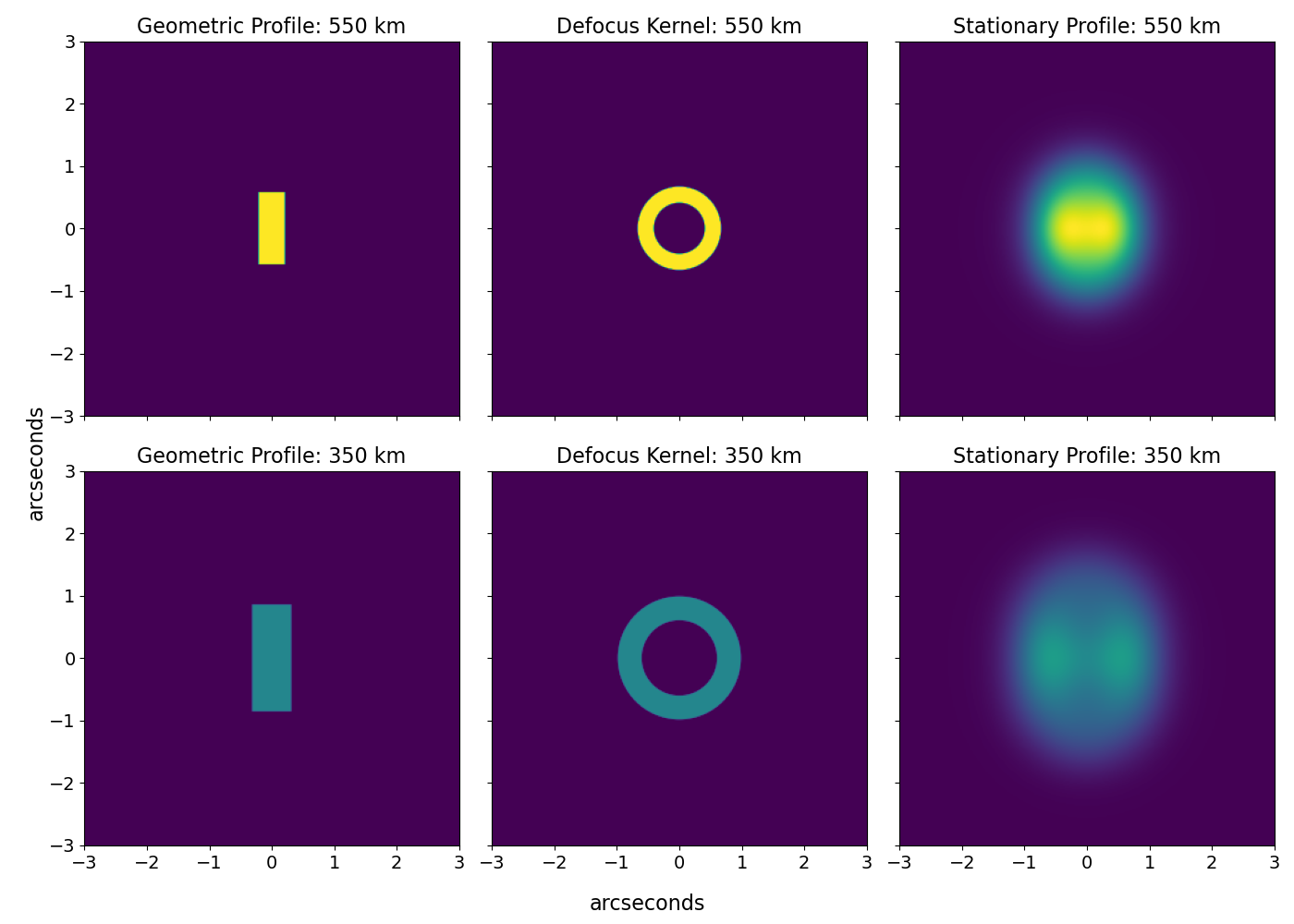}
    \caption{The geometric profiles (left column), defocus kernels (middle column), and stationary surface brightness profiles (right column) of a 7.25 m by 2.65 m rectangular satellite at orbital heights of 550 km (top row) and 350 km (bottom row) observed at a zenith angle of 70 degrees by LSSTCam. The difference in height causes an increase in angular size of the profiles and a decrease in surface brightness.}
    \label{fig:profiles_vs_height}
\end{figure}

We used the open-source Galsim software \citep{galsim2015} to create and convolve the above surface brightness profiles to determine the satellite stationary profile for a satellite at two different orbital heights (550 km and 350 km), observed by LSSTCam at the same zenith angle of 70 degrees. For both cases, the atmospheric PSF was modeled as a Kolmogorov surface brightness profile with a FWHM of 0.67". Figure \ref{fig:profiles_vs_height} shows a comparison of geometric profiles, defocus kernels, and convolved stationary profiles, demonstrating the increase in angular size and a decrease in surface brightness for the satellite at a lower orbital height.

To calculate the true brightness, we define the stationary magnitude of the satellite as the measured magnitude of the satellite if it were to be tracked by the telescope over the entire exposure time. We then set the stationary magnitude of the satellite at a range of 550 km to a fiducial value of 5.0 mags in g-band. This magnitude is then scaled to account for observed angle from zenith, as well as changes in the range and viewing angle due to the differences in orbital height. Using the Rubin Observatory simulation tools, developed as a part of the Rubin Data Management software stack, to account for the satellite SED and telescope throughput, we calculated the total flux for each of the two cases for their appropriate effective pixel exposure time and scale the stationary profiles accordingly. Finally, because the entire profile will scan over a cross-sectional line of pixels during the image exposure time, we sum the flux of the stationary profiles along the direction of travel to obtain the predicted cross-sectional surface brightness profiles.

\begin{figure}[htb]
    \centering
    \includegraphics[width=0.75\linewidth]{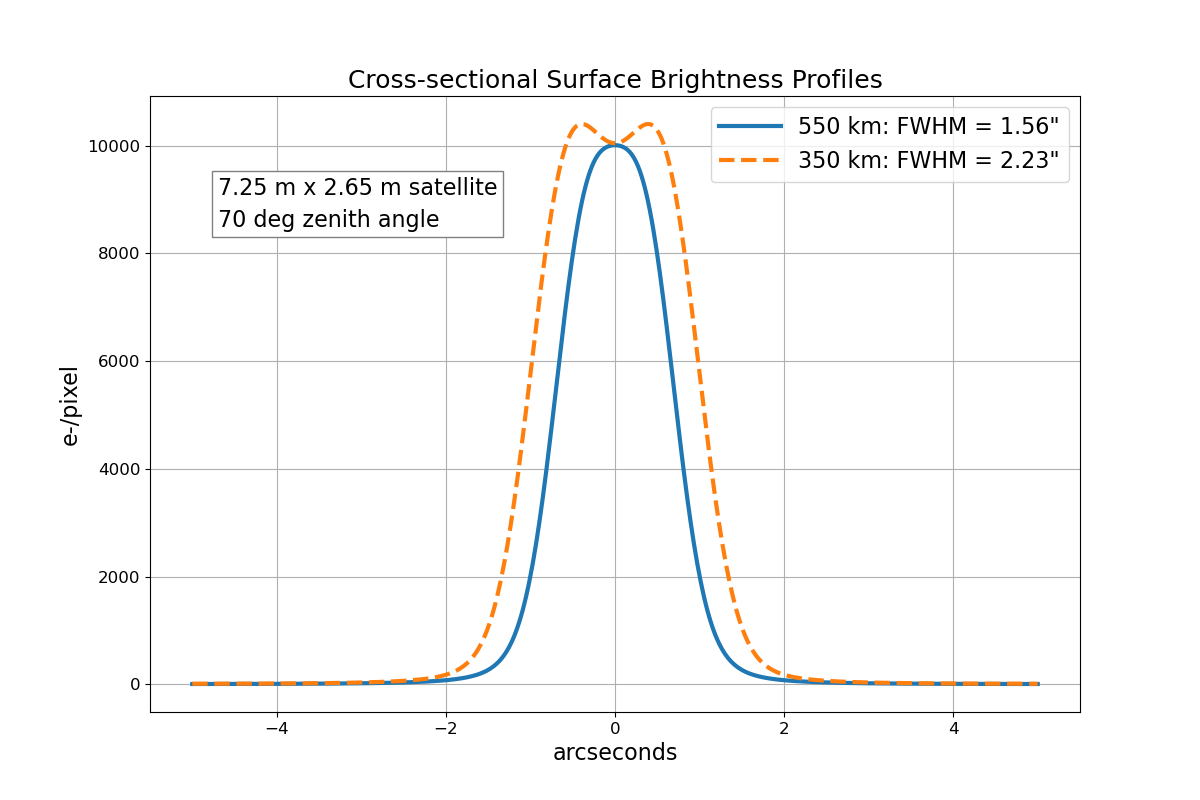}
    \caption{Simulations of the streak cross-sectional surface brightness profiles of a 7.25 m by 2.65 m rectangular satellite at orbital heights of 550 km and 350 km observed at a zenith angle of 70 degrees as it streaks across the focal plane of LSSTCam. This is done for a flux corresponding to a 5th g magnitude tracked satellite, scaled by range and viewing angle. Several factors nearly cancel, giving a peak surface brightness only 4\% higher at 350 km.}
    \label{fig:cross-section_surface_brightness}
\end{figure}

As the satellite surface brightness profile shown in Figure \ref{fig:profiles_vs_height} scans across the LSSTCam focal plane, the observed streak has a profile very different from an in-focus star. For example, the peak surface brightness is typically six times lower than that of a star~\citep{tyson2024expected}. One can also study the effects of lowering the satellite orbit.   Figure \ref{fig:cross-section_surface_brightness} shows the streak surface brightness profiles for satellites at 550 km orbit and 350 km orbit.  Note that the effects due to range, angular velocity, and de-focus nearly cancel out as we move down to 350 km. 

\section{Discussion}\label{sec:discussion}
We have simulated the effect on satellite surface brightness of moving a large partially resolved satellite from an orbit at 550 km down to 350 km. The surprising result is that the surface brightness of the streak on the LSSTCam focal plane is nearly the same. This is true independent of the observed zenith angle of the satellite and of its reflectivity. This result comes from a purely analytic calculation, and does not require reflectivity data for the satellites, since we assume the same unmodified satellite is used in the comparison of optical surface brightness. Thus our result is complementary to satellite brightness prediction analyses which rely on satellite BRDF measurements and on-orbit brightness measurements~\citep{fankhauser23, kandula2025}. Indeed there are additional impacts at lower orbits~\citep{kandula2025}. Finally, we note that there is an increase in the FWHM of the satellite at a lower orbital height; therefore, a wider mask will need to be used resulting in a small increase in the loss of pixel data.

\bigskip
\section*{Software}
The software written for this project is available at a public Github repository: \url{https://github.com/Snyder005/leosim}. Rubin simulation tools are available at the Github repository: \url{https://github.com/lsst/rubin_sim}. \texttt{Lumos-Sat} is an open-source Python package \citep{lumos2023}. Documentation and installation instructions for \texttt{Lumos-Sat} can be found at: \url{https://lumos-sat.readthedocs.io/}. 
Astropy \citep{astropy2022}, 
NumPy \citep{numpy2020}, 
SciPy \citep{scipy2020}, 
Matplotlib \citep{matplotlib2017}, 
pandas \citep{pandas2010}.

\begin{acknowledgments} 
We acknowledge support from NSF grant AST-2205095 and NSF/AURA/LSST grant N56981CC to UC Davis. We acknowledge helpful discussions with Craig Lage, Daniel Polin, and Forrest Fankhauser. 

\end{acknowledgments}

\bibliography{main} 
\bibliographystyle{aasjournalv7} 

\end{document}